\documentstyle[12pt]{article}
\def \pbjpa {\pagebreak}
\newcommand{\prepnumber}[1]{}
\newcommand{\pacs}[1]{
\vspace{10mm}
\begin{flushleft}
PACS.~~#1 \\
\end{flushleft}
\vspace{20mm}}
\def \be {\begin{equation}}
\def \ee {\end{equation}}
\def \baa {\begin{equation}\left.\begin{array}{l}}
\def \eaa {\end{array}\right.\end{equation}}
\def \bea {\begin{equation}\left.\begin{array}{lll}}
\def \eea {\end{array}\right.\end{equation}}
\def \ben {\begin{displaymath}}
\def \een {\end{displaymath}}
\def \ie {{\it i.e. }}

\def \dif {{\rm d}}

\begin{document}

\title{
On the 1/D expansion for directed polymers\\
}
\author{
Giorgio Parisi\\ Dip.  di Fisica, I Universit\`a di Roma ``La Sapienza'' \\
Piazzale Aldo Moro, I-00185~Roma, Italy\\
and INFN, sezione di Roma I\\
 and \\
Frantisek Slanina\\
   Center for Theoretical Study\\
    Jilsk\'a 1, CZ-11000 Praha, Czech 
Republic \\
}
\maketitle
\begin{abstract}
We present a variational approach for directed polymers in $D$ 
transversal dimensions which is used to compute the corrections to the
mean field theory predictions with broken replica symmetry.  The trial
function is taken to be a symmetrized  version of the mean-field solution,
which is known to be exact for  $D=\infty$.  We compute the free energy
corresponding to that function and  show that the finite-$D$ corrections
behave like $D^{-4/3}$.  It means that  the expansion in powers of $1/D$
should be used with great care here. We hope that the techniques
developed in this note will be useful also in the study of spin
glasses.  \end{abstract}

\section{Introduction}

At the present moment a very useful approach in the study of phase 
transitions in disordered systems is given by the mean-field theory in the 
framework of the spontaneous breaking of the replica symmetry 
\cite{MePaVi}.  In many cases there is a well-developed theoretical 
understanding of the mean-field theory, using the replica formalism or 
using the equivalent probabilistic cavity approach.  The situation is not 
so good if we consider the perturbative corrections to the mean field 
approach in the phase where replica symmetry is broken: a few computations 
exist, but they are very difficult from the technical point of view, so 
that only one loop contributions have been computed \cite{DeD-Ko-Te 1991}
\cite{MePa prep}.

The aim of this paper is to explore a new field: non 
perturbative effects. Only very few papers exist which are devoted to the
study of non perturbative effects, among them we recall
\cite{FPV,CC,CAMPELLONE,DM}. Non perturbative effects are very important
because they could spoil the predictions of mean field theory in a rather
subtle way, which cannot be detected by the usual perturbative methods.

 In this
note we will concentrate our attention on  the following problem.  In the phase
where replica symmetry is broken, in  replica space there are many different
stationary points for the free  energy as a function of the order parameter (let
us call it $Q_{a,b}$ and  denote its average in a given stationary point by
$q_{a,b}$).  These stationary points are related to  the other by permutations. 
In the usual approach one pick up a particular  solution and the symmetrization
is done explicitly when one computes the  observables.  For example one writes 
\be <Q_{a,b}^2>={1 \over n!}\sum_\Pi  q_{\Pi(a),\Pi(b)}^2, \ee
where $\Pi$ denotes a permutation, $\Pi(a)$ is value 
of the index $a$ after the permutation and the sum is done over all the 
$n!$ permutations of $n$ elements ($n \to 0$ at the end). 

This symmetrization is crucial when we need to compute quantities in finite 
volume because in this situation replica symmetry must be restored.  The 
physical interpretation of spontaneously broken replica symmetry as the 
coexistence of infinite many pure states is deeply rooted on this 
procedure. For example in
the case of spin glasses we have that
\be
{1\over
N^p}\sum_{i_1,i_2,...,i_p=1,N}<\sigma_{i_1}\sigma_{i_2}...\sigma_{i_p}>^2
\approx <Q_{a,b}^p>={1 \over n!}\sum_\Pi  q_{\Pi(a),\Pi(b)}^2. \ee

Things however become more complicated in systems which have one direction 
much longer that the others.  Let us consider a system with $N L$ degrees 
of freedom (e.g.  $\sigma_{k,t}$), where $k=1,...,N$, $t=1,...,L$) and let 
us suppose that the interaction has a finite range in the variable $t$.  We 
are interested in the case where $L$ is very large (infinite) and $N$ is 
large, but definitely much smaller than $L$.  General arguments tell us
that the correlation length in the $t$ direction must be finite as soon
$N$ is finite: the system becomes one dimensional and in this situation no
spontaneous breaking of a symmetry is possibile.

 For large $N$ we can formally  associate an order parameter to each system at
fixed $t$.  Systems at quite  different $t$ will have order parameters which
point into different  directions in the order parameter space (in our case
replica space); indeed standard arguments imply that also if  we constrain the
order parameter at time zero to point in a given  direction the symmetry will be
restored at large times.  This restoration is related to tunneling events among
configurations in which the order  parameters have different orientations in
replica space.  In this situation  the zero modes connected to the spontaneous
breaking of the replica  symmetry are lifted and a finite correlation length (in
the $t$ direction)  appears in the problem.  Of course this correlation length
diverges in the  limit when $N$ goes to infinity.

This kind of problems appears in many cases, for example if we study the 
dynamics of spin glasses and we are interested in the tunneling among 
different pure states.  A similar problem arises also in the equilibrium 
behavior of samples that are much longer in one direction that the other.

In this paper we study these problems in a different context, the 
equilibrium behavior of infinitely long directed polymer. In the replica
approach one finds that at low temperature the replica symmetry is broken. This
breaking is an artefact of the mean field approximation and replica symmetry is
eventually restored in this problem.  In this case it  is crucial
to find out the solution to the problem of tunneling among  different ground
states, which is likely to control the large time behavior of the correlations.  

The number of 
degrees of freedom of a directed polymer embedded in a $D$-dimensional space is proportional to $D$.  
The replica approach can be successfully used when $D$ goes to infinity, but for fixed $D$ we have 
to deal with the existence of tunneling effects and replica symmetry restoration at large times.  It 
is quite possible that this effects are responsible for the inaccurate value of the exponent 
computed at low dimensions in the replica approach.

In this note we present a tentative computation of tunneling effects in the 
case of breaking of the replica symmetry in a continuous way, i.e. the 
function $q(u)$ is continuous \footnote{In a separate publication we will 
discuss in more details the simpler case where one the replica symmetry is 
broken at one step, i.e.  the function $q(u)$ takes only two values 
\cite{PaSla in prep}.}.  We will show that one can define infinitesimal 
permutations and the tunneling among configurations differing by an 
infinitesimal permutation is not exponentially suppressed and gives a 
contribution that decreases as a power of $D$ at large dimensions.

In the general case the computations are rather involved.  In the present 
case the analysis is simple because we can have to find the ground state of 
the Schroedinger equation for $n$ interacting particles.  The effects of the 
tunneling among wave functions with different order parameter may be taken 
into account in a first approximation by using a symmetrized wave function 
and computing the effects due to the superposition of different wave 
functions.

This paper is organized as follows: in section 2 we recall some of the 
results of the replica approach to directed polymer, we present in the 
next section the symmetrized trial wavefunction and we compute the 
corresponding free energy.  The section 4 is devoted to the discussion of 
the integral over permutations and its consequences to the finite-$D$ 
corrections.  The last section summarizes the results obtained.

\section{Mean-field solution}

The directed polymer in a random external field is described by the Hamiltonian
\cite{KaZha 1987} \cite{Mezard 1990} \cite{MePa 809}
\be
h[\omega]=\int\dif
t\,(\frac{1}{2}\left({\partial\omega\over\partial t}\right)^2+V(t,\omega(t)))
\ee
where $t$ is the coordinate along the polymer and the $D$-component vector 
$\omega$ describes the transversal coordinate.  The random external field 
is supposed to be Gaussian-distributed with the correlation 
\be
\overline{V(t,\omega)V(t^\prime,\omega^\prime)}=-D\delta(t-t^\prime) 
f({|\omega-\omega^\prime|\over\sqrt{D}}) .
\ee
The relevant part of the 
correlation is the long-distance behavior, which we choose to be power-law
\be
f(x)\simeq{g\over 2(1-\gamma)}x^{2(1-\gamma)}\;
,\ x\to\infty\;
.
\ee

In this approach the directed polymer is a classic-mechanical
static object, so that the interesting properties are
related to its shape. Moreover, we neglect the size of the
particular monomers. In this way there is no intrinsic
natural length scale in the system.

The scaling exponents are crucial to describe the large distance
behaviour of the system.  In particular $\zeta$ controls 
 the transversal fluctuations
\be
\overline{\langle(\omega(t)-\omega(t^\prime))^2\rangle}
\sim (t-t^\prime)^{2\zeta}.
\ee
Sometimes $\zeta$ is called wandering exponent. 

The exponent $\chi$ controls  the sample-to-sample fluctuations of the free 
energy
of a finite part of the polymer, of the length $L$
\be
\overline{F(L)^2}-\overline{F(L)}^2\sim L^{2\chi}.
\ee

There is a scaling formula relating $\zeta$ and $\chi$, resulting from the 
Galilean invariance: $\chi=2\zeta-1$ \cite{Huse-Henley 1985} \cite{Medina 
1989}.

Different approaches to the calculation of the scaling exponents are 
possible \cite{MePa 809} \cite{Medina 1989} \cite{Kardar 1987} \cite{Vilgis 
1991}, among them the replica method is one of the most powerful.  Making 
use of the replica trick, the problem is equivalent to the system of $n$ 
particles, where $n$ is the number of replicas, with pairwise interaction 
determined by the correlation function of the random external field.  The 
exact solution is known for two opposite cases.  For $D=1$ and 
$\delta$-correlated disorder the Bethe {\it Ansatz} leads to the value 
$\zeta=\frac{2}{3}$. 

 On the other hand, simple scaling argument (see {\it 
e.g.} \cite{Halpin-Healy 1989}) leads to the Flory result

\be
\zeta=\zeta_F\equiv{3\over 2(1+\gamma)}\; .
\ee

The same scaling analysis gives us the relation for the effective value of 
$\gamma$ for the case of $\delta$-correlated disorder $\gamma=1+D/2$, thus 
the Flory result does not coincide with the exact solution for $D=1$.  The 
question of the validity of the Flory formula arises and it is partially 
answered by the results obtained in another case, where the exact solution 
is known, namely in the $D=\infty$ limit \cite{MePa 809}.  In this case, 
the mean field approach should give the correct result.

The mean-field equations can be obtained in a number of ways. In the most 
transparent method one consider the perturbation expansion for the
propagator  $G_{ab}(t-t^\prime)=\langle\omega_a(t)\omega_b(t^\prime)\rangle$,
where $a$  and $b$ are the replica indices.  The role of the order parameter is
played  by the self-energy $\sigma_{ab}$ defined by
\be
G_{ab}(t)=([-{\partial^2\over\partial
t^2}-\sigma]^{-1})_{ab} \ee

The point is that the Hartree-Fock approximation as exact in the $D=\infty$ 
limit ($D$ corresponds to the number of field components), thus giving the 
desired mean-field equations for $\sigma$. Only cactus diagrams survive in this
limit, thus reproducing the Hartree-Fock expansion. Non-cactus diagrams give
formally corrections which are proportional to $1/D$.

Two cases are possible.
\begin{itemize}
\item  In the situation called `short range 
correlations', the matrix $\sigma$ is either replica-symmetric or it 
corresponds to one step of replica-symmetry breaking.  This solution is 
thermodynamically stable for $\gamma>2$.  The resulting wandering exponent 
is $\zeta=\frac{1}{2}$, {\it i.  e.} no influence of the disorder is seen.  
\item
For $\gamma<2$ another solution should be looked for and 
the natural choice is to break the replica symmetry in the hierarchical 
manner already familiar from the spin-glass theory \cite{MePaVi}.  The 
matrix $\sigma_{ab}$ is then parametrized by the pair 
$(\tilde\sigma,\sigma(u))$, where $\sigma(u)$ is a function on the interval 
(0,1) and the number $\tilde\sigma$ is the diagonal element of the matrix 
$\sigma_{ab}$.  At the end we arrive at a surprising result stating that 
the wandering exponent has exactly the Flory value.
\end{itemize}

 The problem can be formulated in an alternative way by using 
the  Feynman-Katz representation for the solution of the Schroedinger equation. 
In this case, the usual replica approach is  equivalent to finding the ground
state of the imaginary-time quantum  Hamiltonian. 

If we use  the variational
principle and we suppose that the trial  wavefunction ($|\psi>$) is Gaussian,
{\it i.e.} 
\be
\langle\omega|\psi\rangle=  \exp(-Q^{-1}_{ab}\omega_a\omega_b),
\ee
we recover the results of the previously described replica approach.
This approximation corresponds to taking into  account only one of the minima of
the free energy and supposing that this  valley has parabolic shape, \ie it
corresponds to the $D$-dimensional  harmonic oscillator, whose ground state
wavefunction is Gaussian.  The  matrix $Q^{-1}$ of variational parameters is
then equivalent to the  self-energy in the Hartree-Fock approach.

\section{Symmetrized wavefunction}
\subsection{General considerations}
In case where the transverse spatial dimensionality $D$ is finite, the 
trial wavefunction should be generalized in such a way that all the 
equivalent valleys are taken into account.  The situation is analog to the 
double-well potential where the barrier between the valleys are 
proportional to $D$.  For $D$ finite we have to include tunneling from one 
valley to the other one.  The relation between the equivalent valleys is 
provided by the symmetry of the problem, in our case by the permutations of 
the $n$ replicas.  We denote $S_n$ the group of all such permutations and 
$S_n^1$ the subset of $S_n$ obtained by excluding the identity.

The problem we want to solve is to find (in the limit $n \to 0$) the ground state
of the Hamiltonian
\be
H=-\frac{1}{2}\sum_{a=1}^n
({\partial\over\partial\omega_a})^2
+D\sum_{a\neq b}f(\frac{1}{\sqrt{D}}|\omega_a-\omega_b|),
\ee
where the attractive interparticle potential can have two forms
\be
f(x)=-g \delta(x)
\ee
or
\be
\label{f of x}
f(x)={g\over 2(1-\gamma)}x^{2(1-\gamma)}.
\ee
The scaling analysis suggests the first case corresponds to the second case for
$\gamma=1+D/2$. We will use the form (\ref{f of x}) hereafter.

In the usual approach one  minimize the free energy functional
\be
F=\frac{1}{nD}{\langle\psi|H|\psi\rangle
\over \langle\psi|\psi\rangle}
\ee
in the space of Gaussian wave functions.
Here we want to generalize this approach by minimizing the same functional
in the space of the symmetrized wavefunctions of the Gaussian type
\be
|\psi\rangle=\sum_{\Pi\in S_n}\Pi|\psi_1\rangle\label{symm}
\ee\be
\label{psi1}
\langle\omega|\psi_1\rangle=
\exp(-\frac{1}{4}\sum_{a,b}Q_{ab}^{-1}
\omega_a\cdot \omega_b)
\ee
with the matrix $Q$ as a variational parameter. This choice
corresponds to the prescription used in \cite{MePa 809} ,
where the role of $Q^{-1}$ plays the selfenergy, which is
constant in longitudinal momentum space.
The permutation $\Pi$ of the replicas acts as a transformation of the matrices 
$Q$ and $Q^{-1}$: 
\bea
Q_{ab}\to \Pi Q_{ab}\equiv Q_{\Pi(a)\Pi(b)},
Q_{ab}^{-1}\to \Pi Q_{ab}^{-1}\equiv Q_{\Pi(a)\Pi(b)}^{-1}.
\eea 

Unfortunately enough when we try to compute quantities like $\langle\psi|H|\psi\rangle$ or even 
$\langle\psi|\psi\rangle$, we have not been able to write in a closed form the sum over the permutation 
in the case where the matrix $Q$ is such to break explicitly the replica symmetry.  It is quite 
possible that a compact formula for such a sum does exists and can be found in a smart way. 

In this note we have esplored the possibility of selecting those permuations we believe are 
the most relevant are we have done  an approximate evaluation of the 
quantity defined in eq.
\ref{symm}.

\subsection{Infinitesimal permutations}

Let us denote $\delta 
Q^{-1}=\frac{1}{2}(\Pi Q^{-1}-Q^{-1})$ the difference between the permuted 
and the original matrix. Naively speaking we expect that for most of the
permutations, for which $\delta Q^{-1}$ is a quantity of order 1, the
contribution to eq. (\ref{symm}) is very small for large $D$, because the
overlap between the original and permuted wavefunction tends to zero as the
transversal dimensionality $D$ goes to  infinity.  This is certainly true as long
as the number of replicas remains  greater than one.

There are technical difficulties arise when we perform the replica limit $n\to
0$.  One of  the peculiar things happening here is that the Schwartz inequality
is no more valid (the space of the $Q$ has negative dimensionality when $n<1$)
and in many cases  it gets  reversed.  Sometimes one finds that overlap of the
original and permuted vector blows up instead  of vanishing when $D\to\infty$. 
However, we can relate this problem to the  set of strange phenomena in the
replica method such as the change of the  mimima of the free energy into maxima
which arise as an effect of the analytic continuation in $n$.

The common
procedure is to take  the $n\to 0$ limit after all other calculations, which
involves some  changes of order of limits.  In this case we will make the $D\to
\infty$  limit first and the replica limit ($n\to 0$) after that.  More
specifically, we will  compute the integrals of the type $\int \dif x \exp(\alpha
(n) D f(x))$ by  saddle point method in the region $n>1$, where $\alpha $ is
negative and  then we will continue the result to $n<1$, where $\alpha $ is
positive and  strictly speaking the integral is not defined.  However, in fact,
the  integral represents the sum over negative number of terms, thus the 
`undefined' integral should be actually defined by the procedure we use to 
compute it or by another methodologically similar procedure.

If we exclude large permutations, we expect that only
the infinitesimal  permutations, {\it i.e} those for which $\delta Q^{-1}$ is
small, will  contribute. These infinitesimal permutation are not present for
integer $n$ and they arise only as an effect of the analytic continuation in $n$
and of the peculiar way of breaking the replica symmetry.

The infinitesimal permutations we will sum over were already classified by 
Goltsev \cite{Goltsev}.  In his classification the exchanges of two 
replicas play crucial role as generators of the whole group of 
permutations.  Clearly, the pairwise exchanges are not the only 
infinitesimal permutations; one should include the permutations of three, 
four etc.  replicas as well, because we do not refer to the set of 
generators but to the set of permutations with small $\delta Q^{-1}$ in 
general.  Nevertheless, we will use the pairwise exchanges of {\it blocks} 
of replicas as typical representatives of the whole set of infinitesimal 
permutations with the hope that they cover that set densely enough to give 
reasonable results.

We suppose that the matrix $Q^{-1}$ has the usual hierarchical form
\cite{Parisi 1980}
\baa
Q_{aa}^{-1}=\tilde{q}=q(1) \\
Q_{ab}^{-1}=q(x), x\in(0,1)
\eaa
where $x=a\cap b$ is the overlap (ultrametric co-distance) between the 
different replicas $a$ and $b$.  Thus even the matrix $Q$, which is the inverse
of $Q^{-1}$, has the  same hierarchical structure
\baa
Q_{aa}=\tilde{r}=r(1)  \\
Q_{ab}=r(x), x\in(0,1)
\eaa
and explicit relations between the pairs $(\tilde{q},q(x))$ and 
$(\tilde{r},r(x))$ can be found in the appendix of \cite{MePa 809}.  We 
will frequently use these relations in the last part of the calculations.  
We use double notation for the value of the diagonal element for 
convenience in the following calculations.

Let us now define the permutations we will take into account.  We will take 
two blocks of replicas of the size $m_B$ and of the mutual co-distance $m$.  
Clearly $m<m_B$.  The permutation in question, which we will denote 
$\Pi_2(m_B,m)$ will involve the exchange of these blocks as fixed units.  
We will use the notation $S_n^{(2)}$ for the set of all permutations of 
this type, the identity excluded.

Denoting
the chosen blocks by the indices $\alpha$ and $\beta,
\alpha\neq\beta$ and the
single replicas within these blocks by pairs $(\alpha ,i)$
and $(\beta ,j)$ it means that the following relations hold
for the overlaps
\baa
(\alpha ,i)\cap (\beta ,j)=m \;\forall i,j  \\
(\alpha ,i)\cap (\alpha ,j)=(\beta ,i)\cap (\beta ,j)=m_B
\;\forall i,j
\eaa
The explicit expression for the change in the matrix
$Q^{-1}$ is
\bea
\delta Q^{-1}_{(ai)(bj)}=&\frac{1}{2}(
\delta_{a\alpha}\bar{\delta}_{b\alpha}\bar{\delta}_{b\beta}
(Q^{-1}_{(\beta i)(bj)}-Q^{-1}_{(\alpha i)(bj)})+\\
&+\delta_{a\beta}\bar{\delta}_{b\alpha}\bar{\delta}_{b\beta}
(Q^{-1}_{(\alpha i)(bj)}-Q^{-1}_{(\beta i)(bj)})+\\
&+\delta_{b\alpha}\bar{\delta}_{a\alpha}\bar{\delta}_{a\beta}
(Q^{-1}_{(ai)(\beta j)}-Q^{-1}_{(ai)(\alpha j)})+\\
&+\delta_{b\beta}\bar{\delta}_{a\alpha}\bar{\delta}_{a\beta}
(Q^{-1}_{(ai)(\alpha j)}-Q^{-1}_{(ai)(\beta j)}))\; .\\
\eea
We will rewrite the expression for the free energy
\be
\label{free energy}
F=\frac{1}{nD}{\langle\psi_1|H|\psi_1\rangle+\sum_{\Pi\in
S^1_n}\langle\psi_1|H|\psi_{\Pi}\rangle
\over
\langle\psi_1|\psi_1\rangle
+\sum_{\Pi\in
S^1_n}\langle\psi_1|\psi_{\Pi}\rangle}
 \ee
where $|\psi_{\Pi}\rangle=\Pi|\psi_1\rangle$ and
$|\psi_1\rangle$ was defined in (\ref{psi1}). 
We now face the problem of evaluating the various terms in the previous
equation as function of the infinitesimal permutation and eventually to sum over
all the infinitesimal permutations.

\subsection{Some explicit formulae}
Here we evaluate the various terms which appear in eq. (\ref{free energy}) and
we postpone  the sum over the infinitesimal perturbation to the next section.

 For
the overlap between the original and the perturbed wavefunction we find \be
\langle\psi_1|\psi_{\Pi}\rangle=\det\,^{-D/2}(Q^{-1}(1+Q\delta
Q^{-1}))=\langle\psi_1|\psi_1\rangle\exp(-\frac{D}{2}T_{\Pi})
\ee
where the function $T_{\Pi}$ can be expanded in powers of
$\delta Q^{-1}$.
\be
\label{T of Q}
T_{\Pi}={\rm Tr}Q\delta Q^{-1}
-\frac{1}{2}{\rm Tr}Q\delta Q^{-1}Q\delta Q^{-1}+\ldots
\ee

This power expansion will be the starting point of the following 
computations.  Of course, the power of $\delta Q^{-1}$ does not tell us 
what is the order of smallness of the certain term, because the magnitude 
of $ \delta Q^{-1}$ itself depends on the parameters of the permutation in 
question.  Thus, we will proceed by choosing an infinitesimal permutation 
and then collecting the terms of the same order of magnitude, which will 
arise during the computation of the value of the subsequent terms in 
(\ref{T of Q}).

Just the same consideration holds for the matrix elements of
the Hamiltonian between the original and permuted state. The
kinetic part of the Hamiltonian gives
\baa
\label{Hkin}
\langle\psi_{\Pi}|-\frac{1}{2}\sum_a\Delta_a|\psi_1\rangle=\\
\\ \frac{D}{2}[TrQ^{-1}+\frac{1}{2}Tr\delta Q^{-1}Q\delta
Q^{-1}-\frac{1}{2}Tr\delta Q^{-1}Q\delta Q^{-1}Q\delta
Q^{-1}+\ldots]\langle\psi_{\Pi}|\psi_1\rangle
\eaa
and the interaction term
\baa
\label{Hint}
\langle\psi_\Pi|D\sum_{a\neq b}
f(\frac{1}{\sqrt{D}}|\omega_a-\omega_b|)|\psi_1\rangle=\sum_{a\neq
b}\frac{1}{2}\hat{g}D\{ \\ \\
{1\over 1-\gamma}(Q_{aa}+Q_{bb}-2Q_{ab})^{1-\gamma}-\\ \\
(Q_{aa}+Q_{bb}-2Q_{ab})^{-\gamma}([Q\delta Q^{-1}Q]_{aa} +
[Q\delta Q^{-1}Q]_{bb}-2[Q\delta Q^{-1}Q]_{ab})+\\ \\
(Q_{aa}+Q_{bb}-2Q_{ab})^{-\gamma}([Q(\delta Q^{-1}Q)^2]_{aa}
+ [Q(\delta Q^{-1}Q)^2]_{bb}-\\ \\ 2[Q(\delta
Q^{-1}Q)^2]_{ab})+\ldots\}\langle\psi_{\Pi}|\psi_1\rangle
\eaa
The first terms in each of these expansions correspond to
the mean field ($D=\infty$) approach \cite{MePa 1992}.
\be
H_0=F_{D=\infty}=\frac{1}{2}\tilde{q}+\frac{1}{2}\hat{g}
{1\over1-\gamma}\int_0^1\dif x(2(\tilde{r}-r(x)))^{1-\gamma}
\; .
\ee
As we will see in the following, the lowest order of magnitude in the 
smallness of the permutation corresponds to the first and second order in 
$\delta Q^{-1}$, {\it i.e.} the terms up to the order $O((\delta 
Q^{-1})^2)$ are necessary.

The rescaled interaction constant $\hat{g}$ arises in the calculation.  It 
depends on the bare coupling $g$ and the transverse dimensionality $D$.  
For $D\to\infty$ the two coupling constants coincide, $\hat{g}\to g$ 
\cite{MePa 809}.  Because the interesting effects do not regard the 
rescaling of the interaction constant, we chose $\hat{g}$ independent of 
$D$ and $g$ having the corresponding $D$-dependence.

When computing the replica summations involved in (\ref{Hkin}) and 
(\ref{Hint}) we use the tree diagrams which describe the configuration of 
replicas imposed by the ultrametric structure of the replica space.  Let us 
turn to the value of $T_\Pi$ first.  The lowest order in $\delta Q^{-1}$ 
corresponds to two non-zero trees shown on the figure 1a.  The 
corresponding value is then
\be
TrQ\delta Q^{-1}=2m_B\int_m^{m_B}\dif
x(r(x)-r(m))(q(x)-q(m))\; .
\ee
Similarly, the trees contributing to the following term in
the expansion of $T_\Pi$ are depicted on the figure 1b,c and
the value is
\baa
-\frac{1}{2}Tr(Q\delta Q^{-1})^2= -(m_B\int_m^{m_B}\dif x
(r(x)-r(m))(q(x)-q(m))^2+\\ \\
m_B(\tilde{r}-\int_{m_B}^1r(y)\dif y-m_Br(m))\\ \\
{[}\int_m^{m_B}\dif x (\tilde{r}-\int_x^1r(y)\dif y-xr(x))
(q(x)-q(m))^2-\\ \\
\int_m^{m_B}\dif x\int_m^{m_B}\dif y
(r(\min(x,y))-r(m))(q(x)-q(m))(q(y)-q(m))]\; .
\eaa

We will suppose that the function $q(x)$ has the same form
as the one found in the case $D=\infty$, {\it i.e.}
\baa
\label{q of x}
q(x)=Ax^s\;\; x\in (0,x_c)\\
q(x)=Ax_c^s\;\; x\in (x_c,1)\\
\tilde{q}=\mu+\int_0^1q(x)\dif x\; .
\eaa
The parameter $\mu$ was introduced in order to have all
quantities finite. It corresponds to enclosing the system to
a finite volume of the linear dimension $1/\mu$.
Using the relations between $q(x)$ and $r(x)$ listed in
\cite{MePa 809} we find for $m_B<x_c$
\be
T_\Pi=T(m_B,m)=-\frac{1}{3}(s+1)^2(1-{m\over m_B})^3 +
O((1-{m\over m_B})^4)
\ee
while for $m_B>x_c$
\be
T_\Pi=-\frac{m_B}{x_c}(\frac{m_B}{x_c}-1)(1-\frac{m}{x_c})^2
+O((1-\frac{m}{x_c})^3)\; .
\ee
Hence, we see that the values of $m$ which dominate the sum over the 
permutations are those near (and smaller than) $m_B$ for $m_B<x_c$ and 
those near $x_c$ for $m_B>x_c$.  If we suppose the matrix elements of the 
Hamiltonian to have the same power dependence on $1-m/m_B$ and on $1-m/x_c$ 
in the respective regions of $m_B$, the dominating $D$-dependence of the 
free energy comes from the permutations with $m_B<x_c$.  In the following 
we will see that the matrix elements of the Hamiltonian actually do have 
this property.  Thus, the sum over the permutations means an integral over 
the parameters $m, m_B$, $m_B\in(0,x_c)$ and $m\in(0,m_B)$.

Similarly we compute the terms arising in 
$\langle\psi_\Pi|H|\psi_1\rangle$.  Summing of all tree diagrams gives for 
the kinetic part
\baa
\frac{1}{4}Tr(\delta Q^{-1})^2Q\equiv H_1(m_B,m)=\\\\
\frac{m_B}{4}(\tilde{r}-\int_{m_B}^1r(y)\dif y-m_Br(m))
\int_m^{m_B}\dif x (q(x)-q(m))^2 \\\\
+\frac{m_B}{4}[\int_m^{m_B}\dif x(\int_x^1r(y)\dif
y+xr(x)-\tilde{r})(q(x)-q(m))^2+\\\\
\int_m^{m_B}\dif x \int_m^{m_B}\dif y
(r(\min(x,y))-r(m))(q(x)-q(m))(q(y)-q(m)) \; .
\eaa

Using the form of $q(x)$ defined in (\ref{q of x}) we
obtain, for $m_B<x_c$
\be
H_1(m_B,m)=\frac{1}{48}As(s+1)2m_B^{s+1}(1-\frac{m}{m_B})^4
+O((1-\frac{m}{m_B})^5)\; .
\ee

For the interaction terms, similarly
\baa
H_2(m_B,m)=\\
\frac{1}{2}\hat{g}\sum_{a\neq b}
(Q_{aa}+Q_{bb}-2Q_{ab})^{-\gamma}
([Q\delta Q^{-1}Q]_{aa}+[Q\delta Q^{-1}Q]_{bb}-
2[Q\delta Q^{-1}Q]_{ab})=\\ \\
\hat{g}2^{1-\gamma}\{\int_m^1\dif x\int_m^1\dif y
[(\Theta(x,x)-\Theta(x,y))(\tilde{r}-r(\min(x,y)))^{-\gamma}+\\\\
(\Theta(x,x)+\Theta(x,y))(\tilde{r}-r(m))^{-\gamma}]+\\\\
\int_m^1\dif
x[(2\Theta(x,1)-\Theta(x,x)-\Theta(1,1))(\tilde{r}-r(x))^{-\gamma}-\\\\
(2\Theta(x,1)+\Theta(x,x)-\Theta(1,1))(\tilde{r}-r(m))^{-\gamma}]+
2\Theta(1,1)(\tilde{r}-r(m))^{-\gamma}\}
 \eaa
where we have used the auxiliary quantities
\baa
\Theta(x,y)={m_B\over 2}(\Phi_0(x,y)+\Phi_0(y,x)+
\Phi_1(x,y)+\Phi_1(y,x))\\\\
\Phi_0(x,y)=(r(x)-r(m))\int_m^{m_B}\!\dif z\,
(q(x)-q(m)(r(\min(z,y))-r(m))\\\\
\Phi_1(x,y)=\\
\;\;\;=(r(x)-r(m))(q(y)-q(m))(yr(y)+\int_y^1\!\dif
z\,r(z)-\tilde{r})\;\;\;{\rm for}\;y\in(m,m_B)\\
\;\;\;=0\;{\rm otherwise.}
\eaa
In the last term we will write only the lowest order (in $(m_B-m)$ ) terms.

\baa
H_3(m_B,m)=\\\\
-\frac{1}{2}\hat{g}\sum_{a\neq b}
(Q_{aa}+Q_{bb}-2Q_{ab})^{-\gamma}
([Q(\delta Q^{-1}Q)^2]_{aa}
+ [Q(\delta Q^{-1}Q)^2]_{bb}-2[Q(\delta
Q^{-1}Q)^2]_{ab})=\\\\
-\hat{g}2^{1-\gamma}{m_B^2\over 2}\{
\int_m^1\dif x\int_m^1\dif y\,[((r(x)-r(m))^2 -\\\\
(r(x)-r(m))(r(y)-r(m)))(\tilde{r}-r(\min(x,y)))^{-\gamma}+\\\\
((r(x)-r(m))^2+(r(x)-r(m))(r(y)-r(m)))(\tilde{r}-r{m})^{-\gamma}]
+2(\tilde{r}-r(m))^{2-\gamma}+\\\\
\int_m^1\dif
x\,[(\tilde{r}-r(x))^{-\gamma}(2(r(x)-r(m))
(\tilde{r}-r(m))-(r(x)-r(m))^2-(\tilde{r}-r(m))^2)-\\\\
(\tilde{r}-r(m))^{-\gamma}(2(r(x)-r(m))
(\tilde{r}-r(m))+(r(x)-r(m))^2-(\tilde{r}-r(m))^2)]\}\\\\
\int_m^{m_B}\dif x\,(xr(x)+\int_x^1\dif y\,r(y)-\tilde{r})
(q(x)-q(m))^2+O((m_B-m)^4) \; .
\eaa

Substituting the functions $q(x)$ and $r(x)$ we get the following results
\begin{equation}\left.\begin{array}{l}
H_2(m_B,m)=\hat{g}2^{1-\gamma}(-{(s+1)^2\over
As(s+2)})^{-\gamma}\{2{(s+1)^2\over As(s+2)}[({1\over
m_B^{s+2}}-{1\over x_c^{s+2}}+{s+2\over s+1}{1\over
x_c^{s+1}})-\\\\
m_B
({1\over m_B^{s+2}}(1-{s+2\over
s+1}m_B)-{1\over x_c^{s+2}}(1-{s+2\over s+1}x_c))]+
{s+1\over As}(1-\gamma+m_B){1\over m_B^{s+1}}\}\times\\\\
({1\over m_B^{s+2}}-{1\over x_c^{s+2}}+{s+2\over
s+1}{1\over x_c^{s+1}})^{-\gamma}
\frac{1}{3}(s+1)^2(1-{m\over m_B})^3+O((1-m/m_B)^4)
\end{array}\right.\end{equation}

\baa
H_3(m_B,m)=-\hat{g}2^{1-\gamma}{m_B^2\over
2}\{\int_{m_B}^1\dif x[ \phi^2(x)\int_{m_B}^x\dif
y\tilde{\phi}^{-\gamma}(y)+(1-x)\phi^2(x)\tilde{\phi}^{-\gamma}(x)-\\\\
\phi(x)\int_{m_B}^x\dif
y\phi(y)\tilde{\phi}^{-\gamma}(y)-\phi(x)\tilde{\phi}^{-\gamma}(x)
\int_{m_B}^x\dif y\phi(y)]+\\\\
\tilde{\phi}^{-\gamma}(m_B)\int_{m_B}^1\dif
x[(1-m_B)\phi^2(x)+\phi(x)\int_{m_B}^1\dif
y\phi(y)]+2\tilde{\phi}^{2-\gamma}(m_B)+\\\\
\int_{m_B}^1\dif x[(\tilde{\phi}^{-\gamma}(x)-
\tilde{\phi}^{-\gamma}(m_B))(2\phi(x)\tilde{\phi}(m_B)-
\tilde{\phi}^2(m_B))-\\\\
(\tilde{\phi}^{-\gamma}(x)+\tilde{\phi}^{-\gamma}(m_B))
\phi^2(x)]\}
{1\over A}s(s+1)m_B^s\frac{1}{3}(1-{m\over m_B})^3
+O((1-m/m_B)^4)
 \eaa
where
\baa
\phi(x)=r(x)-r(m_B)=-{(s+1)^2\over As(s+2)}({1\over
m_B^{s+2}}-{1\over x^{s+2}})\\\\
\tilde{\phi}(x)=\tilde{r}-r(x)=-{(s+1)^2\over As(s+2)}
({1\over x^{s+2}}-{1\over x_c^{s+2}}+{s+2\over s+1}{1\over
x_c^{s+1}}) \; .
\eaa
\section{Summing over permutations}
Having computed all the matrix elements, we could proceed to computation of 
the free energy (\ref{free energy}).  For this end we first need to know 
the measure $\dif \mu_\Pi$ in the space of permutations.  Starting with the 
discrete $p$-adic formulation of the replica symmetry breaking \cite{PaSou}, where
the  size of the $i$-th block is $m_i=p^i$ and $p^K=n$, we see that there are
\be
\Delta P(m_B,m_i)={n(m_i-m_{i-1})\over 2 m_B^2}
\ee
permutations which exchange blocks of the size $m_B$ at the co-distance 
$m_i$.  The continuum limit corresponds to taking $p\to 1^-$, while 
$K\to\infty$ and $n=p^K\to 0$.  Note that the usual replica limit $n\to 0$ 
and the limit which introduces the continuous replica symmetry breaking, 
$p\to 1^-$ should be taken independently, as in the spin-glass theory 
\cite{MePaVi}.  Thus, the measure in the set $S_n^{(2)}$ of the 
permutations in question is
\be
\dif\mu^{(2)}(m_B,m)=-{n\dif m_B\dif m\over 2 \ln p \, m_B^3}
\ee
and it is immediately seen to diverge for $p\to 1^-$.

The meaning of the divergence clarifies itself once we take into account 
the rest of the set $S_n^1$ of all possible non-identical permutations.  In 
the limit $n\to 0$ $|S_n^1|=n!-1=0$ and the infinite number of permutations 
$\in S_n^{(2)}$ should be canceled by the contribution of other 
permutations.  Actually, if we find {\it e.g.} the measure corresponding to 
the space $S_n^{(3)}$ of all cyclic exchanges of three blocks of replicas, 
we see that their number is infinite again, but with opposite sign than the 
number of permutations in $S_n^{(2)}$.  It is that cancellation of 
infinities that makes finally the total number of non-identical 
permutations to be zero.

It would be very difficult task to perform the same calculations as those 
presented above for all the variety of possible permutations and to show 
the cancellation of the divergences explicitly.  Instead, we will 
approximate the integral over all permutations with the exact measure by 
the much more simple integral over the set $S_n^{(2)}$ only, but taking a 
``renormalized'' measure $\dif\bar{\mu}^{(2)}(m_B,m)$ instead of the true 
$\dif\mu^{(2)}(m_B,m)$.  The interpretation we attribute to such a 
replacement is that we pick out the permutation $\Pi$ with given $m_B$ and 
$m$ and suppose that the other permutations $\bar\Pi$, exchanging not only 
two, but three, four etc.  blocks of replicas are close to $\Pi$ in the 
sense that corresponding $T_\Pi$ and the matrix elements of the Hamiltonian 
are approximately equal, but when summing over all $\bar\Pi$'s, the 
divergences cancel, yielding the average measure 
$\dif\bar{\mu}^{(2)}(m_B,m)$ which is free of the divergence at $p\to 1$.  
Clearly, the above supposition is somewhat unfounded and its mathematical 
formulation is worth of further study.  Nevertheless we consider the 
hypothesis to be plausible and useful at least for the purpose of 
establishing the power of the finite size corrections, which is what we are 
trying here.  Thus, we will rely on it in the following.

The question then arises, how to choose the approximate measure 
$\dif\bar{\mu}^{(2)}(m_B,m)$.  The measure has to meet the condition
\be
\int_{m\in(0,m_B)}\dif\bar{\mu}^{(2)}(m_B,m)=\dif\mu^{(1)}(m_B)
\ee
where $\dif\mu^{(1)}(m_B)$ is the number of all permutations of block of 
the size between $m_B$ and $m_B+\dif m_B$ with the restriction that they 
cannot be expressed as a permutation with another (smaller) block size.  
This restriction avoids multiple counting of the permutations.  Clearly, 
the sum over the block sizes from 1 up to given $m_B$ should give 
$(n/m_B)!$ , so
\be
\dif\mu^{(1)}(m_B)={\dif\over\dif m_B}\Gamma({n\over
m_B}+1)\dif m_B
\stackrel{n\to 0}{=}-\Gamma^\prime(1){n\over m_B^2}\dif
m_B\;
. \ee
Similarly, the following relation holds for the original
measure $\mu^{(2)}$
\be
\int_{m\in(0,m_B)}\dif\mu^{(2)}(m_B,m)=-{n\over 2 \ln p\,
m_B^2} \dif m_B
\ee
hence, {\it per analogiam} we infer that
\be
\dif\bar{\mu}^{(2)}(m_B,m)=-\Gamma^\prime(1){n\over
m_B^3}\dif m\dif m_B\; .
\ee

The same result comes out also following another consideration.  Dividing 
all the permutations into disjunct groups according their characteristics 
$m_B$ (by the same means as we did when we were deriving $\mu^{(1)}$) and 
within a selected group taking only such a permutations which exchange two 
blocks, we obtain exactly the same result.  This approach is based on the 
tacit assumption, that the pair exchanges dominate all quantities of 
interest.  Actually, when we perform the calculation of 
$T_\Pi=T(m_B,m_1,m_2)$ in the case of a cyclic exchange of three blocks 
(the permutation is characterized by two co-distances $m_1<m_2$) we find 
that $T(m_B,m_1,m_2)-T(m_B,m_1)\sim(1-(m_2/m_B))^3>0$ and for large $D$ the 
pair exchanges actually dominate.

The sentence just stated is slightly incorrect in that it holds only for 
$n>1$.  In the replica limit $n\to 0$ the inequality reverses and one would 
naively expect that, on the contrary, the pair exchanges are negligible.  
But the same consideration as the one which has clarified the wrong sign in 
the exponential $\exp(-DT_\Pi/2)$ applies here and the previous statement 
about the dominating permutations is restored.

Finally, we obtain for the free energy
\be
F=H_0-\Gamma^\prime(1)\int{\dif m\dif m_B\over m_B^3}
H(m_B,m)\exp(-{D\over
2}T(m_B,m))
\ee
where $H=H_1+H_2+H_3$.  Note that the kinetic term, $H_1$ is of higher 
order in $(m_B-m)$ than the interaction one, $H_2+H_3$, so the kinetic term 
can be omitted completely in the computation of the finite-dimensionality 
corrections.

Setting $\bar{m}=1-m/m_B$ we can see that the free energy
has the form
\be
F-H_0=\int\dif m_B C(m_B)\int_0^1\dif\bar{m}\,\bar{m}^3
{\rm e}^{-\frac{D}{2}(-\frac{1}{3})(s+1)^2\bar{m}^3}
\ee
and the integral over $\bar{m}$ can be easily computed, yielding
\be
\int_0^1\dif\bar{m}\,\bar{m}^3
{\rm e}^{-\frac{D}{2}(-\frac{1}{3})(s+1)^2\bar{m}^3}
\simeq 
-3^{5/3} 2^{4/3}\Gamma (\frac {4}{3})(s+1)^{-8/3} D^{-4/3} ,(D\to\infty);
.  \ee

The main inference from this result is the $D$-dependence of
the correction to the free energy, which is
\be
F-H_0\sim D^{-4/3} .
\ee
Of course, the full solution should continue with minimizing the free 
energy with respect of the matrix $Q^{-1}$, but the non-integer power of 
$D$ in the corrections to free energy and propagator must persist.  The 
open question and the most interesting one is, however, what are the 
finite-$D$ corrections to the exponent $s$, which is the quantity which 
governs the long-distance behavior of a directed polymer.  In answering 
this question, we cannot avoid the explicit computation of the stationary 
point of the free energy.  In principle it is straightforward: we take the 
formulae just obtained and let the first derivatives with respect to the 
three variational parameters $s, A, x_c$ be zero.  This work is in progress 
and we will refer the reader to further publication.

\section{Conclusions}

We have seen that infinitesimal permutations can be identified in the case 
of continuous breaking of the replica symmetry.  The sum over these 
infinitesimal permutations is not a simple task: if one proceeds in a naive 
way the measure over a class of infinitesimal permutations is formally 
infinite and this infinity is related to having limited our attention on a 
restricted class of permutations.  The corresponding double counting 
problem gives rise to this divergence.  We tentatively propose to cure this 
divergence by introducing an effective measure in order to take care of 
these double counting problems.

If we follow this tentative approach we can compute the shift in the ground 
state energy due to the presence of the tunneling and we find that in a 
given range of the parameters there are corrections which go to zero like 
$D^{- 4/3}$.  The effects of these corrections on the critical exponents 
has not been studied yet.  However, we think that the tunneling among 
different states plays a crucial role in the final theory.  It would be 
extremely interesting to compare these results with those coming from a 
standard $1/D$ expansion to see if these effects are correctly taken 
account in the perturbative $1/D$ expansion or they have a more non 
perturbative nature, as this study may suggest.
\paragraph{Acknowledgments}
F.S. wishes to thank to the University of Rome ``Tor
Vergata'' for the financial support.
\pbjpa
\pagebreak
%
%
%
\pagestyle{empty}
\topmargin=0cm \textheight=25cm \textwidth=16.9cm 
\oddsidemargin=-.3cm
\setlength{\unitlength}{0.32in}
\LARGE
\begin{eqnarray*}
\begin{picture}(7,9)(-1,-1)
\put(0,1){\line(1,3){2}}
\put(4,1){\line(-1,3){2}}
\put(3,1){\line(-1,3){1.5}}
\put(0,1){\line(0,-1){1}}
\put(3,1){\line(0,-1){1}}
\put(-1,1){\vector(1,0){1}}
\put(0.5,5.5){\vector(1,0){1}}
\put(1,7){\vector(1,0){1}}
\put(-1,0.95){\makebox(0,0)[r]{$m_B$}}
\put(0.2,5.45){\makebox(0,0){$u$}}
\put(0.7,6.95){\makebox(0,0){$m$}}
\put(4.4,0.95){\makebox(0,0){$\beta$}}
\put(0.4,0.95){\makebox(0,0){$\alpha$}}
\put(2.9,0.95){\makebox(0,0)[r]{$b$}}
\put(0,-0.1){\makebox(0,0)[t]{$i$}}
\put(3,-0.1){\makebox(0,0)[t]{$j$}}
\end{picture}
\begin{picture}(1,9)(0,-1)
\put(0.5,3.5){\makebox(0,0){+}}
\end{picture}
\begin{picture}(7,9)(-1,-1)
\put(0,1){\line(1,3){2}}
\put(4,1){\line(-1,3){2}}
\put(1,1){\line(1,3){1.5}}
\put(0,1){\line(0,-1){1}}
\put(1,1){\line(0,-1){1}}
\put(5,1){\vector(-1,0){1}}
\put(3.5,5.5){\vector(-1,0){1}}
\put(3,7){\vector(-1,0){1}}
\put(5,0.95){\makebox(0,0)[l]{$m_B$}}
\put(3.5,5.45){\makebox(0,0)[l]{$u$}}
\put(3,6.95){\makebox(0,0)[l]{$m$}}
\put(3.6,0.95){\makebox(0,0){$\beta$}}
\put(0.4,0.95){\makebox(0,0){$\alpha$}}
\put(1.1,0.95){\makebox(0,0)[l]{$b$}}
\put(0,-0.1){\makebox(0,0)[t]{$i$}}
\put(1,-0.1){\makebox(0,0)[t]{$j$}}
\end{picture}
\end{eqnarray*}
\begin{center}
Figure 1a
\end{center}
%
%
\setlength{\unitlength}{0.32in}
\begin{eqnarray*}
\left(
\begin{picture}(7,5)(-2,3)
\put(0,1){\line(1,3){2}}
\put(4,1){\line(-1,3){2}}
\put(3,1){\line(-1,3){1.5}}
\put(0,1){\line(0,-1){1}}
\put(3,1){\line(0,-1){1}}
\put(-1,1){\vector(1,0){1}}
\put(0.5,5.5){\vector(1,0){1}}
\put(1,7){\vector(1,0){1}}
\put(-1,0.95){\makebox(0,0)[r]{$m_B$}}
\put(0.2,5.45){\makebox(0,0){$u$}}
\put(0.7,6.95){\makebox(0,0){$m$}}
\put(4.4,0.95){\makebox(0,0){$\beta$}}
\put(0.4,0.95){\makebox(0,0){$\alpha$}}
\put(2.9,0.95){\makebox(0,0)[r]{$b$}}
\put(0,-0.1){\makebox(0,0)[t]{$i$}}
\put(3,-0.1){\makebox(0,0)[t]{$j$}}
\end{picture}
\begin{picture}(1,5)(0,3)
\put(0.5,3.5){\makebox(0,0){+}}
\end{picture}
\begin{picture}(7,5)(0,3)
\put(0,1){\line(1,3){2}}
\put(4,1){\line(-1,3){2}}
\put(1,1){\line(1,3){1.5}}
\put(0,1){\line(0,-1){1}}
\put(1,1){\line(0,-1){1}}
\put(5,1){\vector(-1,0){1}}
\put(3.5,5.5){\vector(-1,0){1}}
\put(3,7){\vector(-1,0){1}}
\put(5,0.95){\makebox(0,0)[l]{$m_B$}}
\put(3.5,5.45){\makebox(0,0)[l]{$u$}}
\put(3,6.95){\makebox(0,0)[l]{$m$}}
\put(3.6,0.95){\makebox(0,0){$\beta$}}
\put(0.4,0.95){\makebox(0,0){$\alpha$}}
\put(1.1,0.95){\makebox(0,0)[l]{$b$}}
\put(0,-0.1){\makebox(0,0)[t]{$i$}}
\put(1,-0.1){\makebox(0,0)[t]{$j$}}
\end{picture}
\right)^2
\end{eqnarray*}
\begin{center}
Figure 1b
\end{center}
\pagebreak
%
%
%
\topmargin=0cm \textheight=25cm \textwidth=19.9cm 
\oddsidemargin=-1.3cm
\large
\setlength{\unitlength}{0.20in}
\begin{eqnarray*}
\left(
\begin{picture}(3,2)
\put(0,0){\line(1,3){1}}
\put(2,0){\line(-1,3){1}}
\put(2,3){\vector(-1,0){1}}
\put(2,3){\makebox(0,0)[l]{$m_B$}}
\put(0,-0.1){\makebox(0,0)[t]{$j$}}
\put(2,-0.1){\makebox(0,0)[t]{$k$}}
\end{picture}
\right)
\times
\left(
\begin{picture}(7,5)(-2,3)
\put(0,1){\line(1,3){2}}
\put(4,1){\line(-1,3){2}}
\put(3,1){\line(-1,3){1.5}}
\put(2,1){\line(-1,3){1}}
\put(3,1){\line(0,-1){1}}
\put(2,1){\line(0,-1){1}}
\put(-1,1){\vector(1,0){1}}
\put(0,4){\vector(1,0){1}}
\put(0.5,5.5){\vector(1,0){1}}
\put(1,7){\vector(1,0){1}}
\put(-1,0.95){\makebox(0,0)[r]{$m_B$}}
\put(0,3.95){\makebox(0,0)[r]{$u$}}
\put(0.5,5.45){\makebox(0,0)[r]{$v$}}
\put(0.7,6.95){\makebox(0,0){$m$}}
\put(4.4,0.95){\makebox(0,0){$\beta$}}
\put(0.4,0.95){\makebox(0,0){$\alpha$}}
\put(1.9,0.95){\makebox(0,0)[r]{$a$}}
\put(2.9,0.95){\makebox(0,0)[r]{$d$}}
\put(2,-0.1){\makebox(0,0)[t]{$i$}}
\put(3,-0.1){\makebox(0,0)[t]{$l$}}
\end{picture}
\begin{picture}(2,5)(0,3)
\put(0,3.5){\makebox(0,0){+}}
\end{picture}
\begin{picture}(7,5)(0,3)
\put(0,1){\line(1,3){2}}
\put(4,1){\line(-1,3){2}}
\put(3,1){\line(-1,3){1.5}}
\put(2,1){\line(-1,3){1}}
\put(3,1){\line(0,-1){1}}
\put(2,1){\line(0,-1){1}}
\put(5,1){\vector(-1,0){1}}
\put(0.5,5.5){\vector(1,0){1}}
\put(0,4){\vector(1,0){1}}
\put(3,7){\vector(-1,0){1}}
\put(5,0.95){\makebox(0,0)[l]{$m_B$}}
\put(0,3.95){\makebox(0,0)[r]{$v$}}
\put(0.5,5.45){\makebox(0,0)[r]{$u$}}
\put(3,6.95){\makebox(0,0)[l]{$m$}}
\put(3.6,0.95){\makebox(0,0){$\beta$}}
\put(0.4,0.95){\makebox(0,0){$\alpha$}}
\put(2.9,0.95){\makebox(0,0)[r]{$a$}}
\put(1.9,0.95){\makebox(0,0)[r]{$d$}}
\put(2,-0.1){\makebox(0,0)[t]{$l$}}
\put(3,-0.1){\makebox(0,0)[t]{$i$}}
\end{picture}
\begin{picture}(2,5)(0,3)
\put(0,3.5){\makebox(0,0){+}}
\end{picture}
\begin{picture}(7,5)(0,3)
\put(0,1){\line(1,3){2}}
\put(4,1){\line(-1,3){2}}
\put(1.5,1){\line(-1,3){0.75}}
\put(2.5,1){\line(1,3){0.75}}
\put(1.5,1){\line(0,-1){1}}
\put(2.5,1){\line(0,-1){1}}
\put(5,1){\vector(-1,0){1}}
\put(-0.25,3.25){\vector(1,0){1}}
\put(4.25,3.25){\vector(-1,0){1}}
\put(3,7){\vector(-1,0){1}}
\put(5,0.95){\makebox(0,0)[l]{$m_B$}}
\put(4.25,3.2){\makebox(0,0)[l]{$v$}}
\put(-0.25,3.2){\makebox(0,0)[r]{$u$}}
\put(3,6.95){\makebox(0,0)[l]{$m$}}
\put(3.6,0.95){\makebox(0,0){$\beta$}}
\put(0.4,0.95){\makebox(0,0){$\alpha$}}
\put(1.4,0.95){\makebox(0,0)[r]{$a$}}
\put(2.6,0.95){\makebox(0,0)[l]{$d$}}
\put(2.5,-0.1){\makebox(0,0)[t]{$l$}}
\put(1.5,-0.1){\makebox(0,0)[t]{$i$}}
\end{picture}
\begin{picture}(1,5)(0,3)
\put(0,3.5){\makebox(0,0){+}}
\end{picture}
\right.
\\
\left.
\begin{picture}(2,5)(0,3)
\put(0,3.5){\makebox(0,0){+}}
\end{picture}
\begin{picture}(7,5)(0,3)
\put(0,1){\line(1,3){2}}
\put(4,1){\line(-1,3){2}}
\put(3,1){\line(-1,3){1.5}}
\put(2,1){\line(1,3){0.5}}
\put(3,1){\line(0,-1){1}}
\put(2,1){\line(0,-1){1}}
\put(5,1){\vector(-1,0){1}}
\put(0.5,5.5){\vector(1,0){1}}
\put(1.5,2.5){\vector(1,0){1}}
\put(3,7){\vector(-1,0){1}}
\put(5,0.95){\makebox(0,0)[l]{$m_B$}}
\put(1.5,2.45){\makebox(0,0)[r]{$v$}}
\put(0.5,5.45){\makebox(0,0)[r]{$u$}}
\put(3,6.95){\makebox(0,0)[l]{$m$}}
\put(3.6,0.95){\makebox(0,0){$\beta$}}
\put(0.4,0.95){\makebox(0,0){$\alpha$}}
\put(1.9,0.95){\makebox(0,0)[r]{$a$}}
\put(2.9,0.95){\makebox(0,0)[r]{$d$}}
\put(2,-0.1){\makebox(0,0)[t]{$i$}}
\put(3,-0.1){\makebox(0,0)[t]{$l$}}
\end{picture}
\begin{picture}(2,5)(0,3)
\put(0,3.5){\makebox(0,0){+}}
\end{picture}
\begin{picture}(7,5)(0,3)
\put(0,1){\line(1,3){2}}
\put(4,1){\line(-1,3){2}}
\put(2,1){\line(-1,3){1}}
\put(3,1){\line(-2,3){2}}
\put(3,1){\line(0,-1){1}}
\put(2,1){\line(0,-1){1}}
\put(5,1){\vector(-1,0){1}}
\put(0,4){\vector(1,0){1}}
\put(3,7){\vector(-1,0){1}}
\put(5,0.95){\makebox(0,0)[l]{$m_B$}}
\put(0,3.95){\makebox(0,0)[r]{$u$}}
\put(3,6.95){\makebox(0,0)[l]{$m$}}
\put(3.6,0.95){\makebox(0,0){$\beta$}}
\put(0.4,0.95){\makebox(0,0){$\alpha$}}
\put(1.9,0.95){\makebox(0,0)[r]{$a$}}
\put(2.9,0.95){\makebox(0,0)[r]{$d$}}
\put(2,-0.1){\makebox(0,0)[t]{$i$}}
\put(3,-0.1){\makebox(0,0)[t]{$l$}}
\end{picture}
\begin{picture}(2,5)(0,3)
\put(0,3.5){\makebox(0,0){+}}
\end{picture}
\begin{picture}(7,5)(0,3)
\put(0,1){\line(1,3){2}}
\put(4,1){\line(-1,3){2}}
\put(3,1){\line(-1,3){1.5}}
\put(3,1){\line(0,-1){1}}
\put(3,0){\circle*{0.2}}
\put(5,1){\vector(-1,0){1}}
\put(0.5,5.5){\vector(1,0){1}}
\put(3,7){\vector(-1,0){1}}
\put(5,0.95){\makebox(0,0)[l]{$m_B$}}
\put(0.5,5.45){\makebox(0,0)[r]{$u$}}
\put(3,6.95){\makebox(0,0)[l]{$m$}}
\put(3.6,0.95){\makebox(0,0){$\beta$}}
\put(0.4,0.95){\makebox(0,0){$\alpha$}}
\put(3,-0.1){\makebox(0,0)[t]{$(ai)(dl)$}}
\end{picture}
\begin{picture}(7,5)(0,3)
\put(1,3.5){\makebox(0,0){+ ($\alpha$ and $\beta$ exchanged)}}
\end{picture}
\right)
\end{eqnarray*}
\begin{center}
\LARGE
Figure 1c
\end{center}


\vskip4cm
\section*{Figure caption}
\bf Fig.  1 a, b and c. \rm Tree diagrams corresponding to the
summations
over replica indices in the equation (\ref{T of Q}).

\begin{thebibliography}{xx}
%
\bibitem{MePaVi}M. M\'ezard, G. Parisi, M. Virasoro, Spin
glass theory and beyond , World Scientific, Singapore, 1987.
%
\bibitem{DeD-Ko-Te 1991}C. De Dominicis, I. Kondor and T.
Temesv\'ari, {\it J. Phys. A: Math. Gen. } {\bf 24 } \rm (1991)
L301-L308.
%
\bibitem{MePa prep}M. M\'ezard and G. Parisi,
{\it J. Phys I France }{\bf 2 } (1992) 2231 .
%
\bibitem{FPV} J.~Kurchan, G.~Parisi, and M.~A. Virasoro, J. Phys. I France {\bf 3}, 1819 (1993).

\bibitem{CC} D.  Carlucci, {\it On the large order behaviour of the Potts model} cond-mat/9605040, S.  
Caracciolo and D.  Carlucci, (unpublished).

\bibitem{CAMPELLONE} M.  Campellone, {\it Some non perturbative calculations on spin glasses},
cond-mat/9410107.

\bibitem{DM} V.  Dotsenko and M.  M\'ezard {\it Vector breaking of replica symmetry in some low
temperature disordered systems} cond-mat/9611017.
\
\bibitem{PaSla in prep} G.  Parisi and S.  Slanina (in preparation).
%
\bibitem{KaZha 1987}M. Kardar and Y.-C. Zhang, {\it Phys.
Rev. Lett. }{\bf 58 } \rm (1987) 2087.
%
\bibitem{Mezard 1990}M. M\'ezard {\it J. Phys. France }{\bf 51 }
\rm (1990) 1831.
%
\bibitem{MePa 809}M. M\'ezard and G. Parisi,
{\it J. Phys I France }{\bf 1 }\rm (1991) 809.
%
\bibitem{Huse-Henley 1985}D.A. Huse and C.L. Henley, {\it
Phys. Rev. Lett. }{\bf 54 }(1985) \rm 2708.
%
\bibitem{Medina 1989}E. Medina, T. Hwa, M. Kardar and Y.-C.
Zhang {\it Phys. Rev. }{\bf A39 }\rm (1989) 3053.
%
\bibitem{Kardar 1987}M. Kardar {\it Nucl. Phys. }{\bf B290 }\rm
(1987) 582.
%
\bibitem{Vilgis 1991}T.A. Vilgis {\it J. Phys. I France }{\bf 1 }
\rm (1991) 1389.
%
\bibitem{Halpin-Healy 1989}T. Halpin-Healy {\it Phys. Rev.
Lett. }{\bf 62 }\rm (1989) 442.
%
\bibitem{Goltsev}A. V. Goltsev, {\it J. Phys.
A: Math. Gen. }{\bf 24 }\rm (1991) 307.
%
\bibitem{Parisi 1980}G. Parisi, {\it J. Phys. A: Math. Gen. }
{\bf 13 }\rm (1980) L115.
%
\bibitem{MePa 1992}M. M\'ezard and G. Parisi, {\it J. .Phys.
A: Math. Gen. }{\bf 25 }\rm (1992) 4521.

\bibitem{PaSou}G. Parisi and N. Sourlas, in preparation.
%
\end{thebibliography}
\end{document}